\def\be{\begin{equation}}
\def\ee{\end{equation}}
\def\ba{\begin{array}}
\def\ea{\end{array}}

\documentclass[aps,amsmath,amssymb,amsfonts,pra]{revtex4}
\usepackage{graphicx}
\usepackage{epstopdf}
\def\qed{\leavevmode\unskip\penalty9999 \hbox{}\nobreak\hfill
     \quad\hbox{\leavevmode  \hbox to.77778em{%
               \hfil\vrule   \vbox to.675em%
               {\hrule width.6em\vfil\hrule}\vrule\hfil}}
     \par\vskip3pt}

\newtheorem{theorem}{Theorem}
\newtheorem{corollary}{Corollary}

\begin{document}

\title{Monogamy of entanglement for tripartite systems}

\author{Xue-Na Zhu$^{1}$}
\author{Gui Bao$^{1}$}
\author{Zhi-Xiang Jin$^{2}$}
\author{Shao-Ming Fei$^{3}$}

\affiliation{$^1$School of Mathematics and Statistics Science, Ludong University, Yantai 264025, China\\
$^2$School of Computer Science and Technology, Dongguan University of Technology, Dongguan, 523808, China\\
$^3$School of Mathematical Sciences, Capital Normal University, Beijing 100048, China}

\begin{abstract}
We study the monogamy of arbitrary quantum entanglement measures $E$ for tripartite quantum systems. Both sufficient and necessary conditions for $E$ to be monogamous in terms of the $\alpha$th power of $E$ are explicitly derived. It is shown that such monogamy of a entanglement measure $E$ only depends on the boundedness of the solution set of certain equations. Moreover, the monogamy conditions have been also obtained with respect to certain subsets of quantum states for a given quantum correlation. Detailed examples are given to illustrate our results.
\end{abstract}

\maketitle

\section{Introduction}

Quantum entanglement \cite{Q1} among the subsystems of a multipartite system play significant roles in many information processing tasks. A fundamental property of quantum entanglement is the monogamy. The monogamy relations give rise to the distributions of quantum entanglement in the multipartite setting. Monogamy is also an essential feature allowing for security in
quantum key distribution. For a tripartite system $A$, $B$ and $C$ associated with finite dimensional Hilbert spaces $H_A$, $H_B$ and $H_C$, respectively, the monogamy of an quantum entanglement measure $E$ implies that the correlation $E(\rho_{A|BC})$ between $A$ and $BC$ satisfies \cite{m1}
\begin{equation}\label{m11}
E(\rho_{A|BC})\geq E(\rho_{AB})+E(\rho_{AC})
\end{equation}
for any $\rho_{ABC}\in H_A\otimes H_B\otimes H_C$, where $\rho_{AB}=Tr_C(\rho_{ABC})$ and $\rho_{AC}=Tr_B(\rho_{ABC})$ are the corresponding reduced bipartite states. An important issue is to determine whether a given entanglement measure $E$ is monogamous or not. Considerable efforts have been devoted
to this problem \cite{E1,E2,E3,E4,PRL.117.060501,G1,G2,G3,JIn} for quantum correlations
such as entanglement of formation (EOF) \cite{E3}, concurrence \cite{C1,C2,C3}, negativity \cite{N1,N2,N3} and concurrence of assistance \cite{zhu2}.

If $E$ does not satisfy the relation (\ref{m11}), it is still possible to find a positive power $\alpha\in(0,+\infty)$ such that $E^{\alpha}$ satisfies the relation (\ref{m11})
for $\alpha>0$ \cite{G1},
\begin{equation}\label{m12}
E^{\alpha}(\rho_{A|BC})\geq E^{\alpha}(\rho_{AB})+E^{\alpha}(\rho_{AC}).
\end{equation}
The $\alpha$th $(\alpha\geq2)$ power of concurrence and the $\alpha$th $(\alpha\geq\sqrt{2})$ power entanglement of formation for $n$-qubit states satisfy the relation (\ref{m12})
\cite{zhu1}.

In Ref. \cite{PRL.117.060501} the authors show that $E$ is
monogamous if there exists a nontrivial continuous function $f$
such that the following generalized monogamy relation
\begin{equation}
E(\rho_{A|BC})\geq f(E(\rho_{AB}),E(\rho_{AC})),
\end{equation}
is satisfied for states $\rho_{ABC}$. In Ref. \cite{JIn} the authors presented a new parameterized monogamy relation of entanglement with equality,
\begin{equation}\label{jine}
E(\rho_{A|BC})=\mu E(\rho_{AB})+E(\rho_{AC}),
\end{equation}
where $\mu>0$ and $E(\rho_{AB})\leq E(\rho_{AC})$. Monogamy relations satisfied by
the $\alpha$th power of entanglement measures are presented based on Eq. (\ref{jine}) Ref. \cite{JIn}.

In this paper, we investigate the monogamy properties of arbitrary quantum entanglement measures for tripartite quantum systems. We derive explicitly both the sufficient and necessary conditions for $E$ to be monogamous in terms of the $\alpha$th power of $E$, with respect to either all the quantum states. And, the monogamy conditions have been also obtained with respect to certain subsets of quantum states for quantum correlations.

\section{ Monogamy relations for tripartite systems}

In the following we say a quantum  entanglement measure $E$ is $\alpha$-monogamous if there exists a real number $\alpha>0$ such that $E$ satisfies (\ref{m12}), and
$E$ is non-monogamous if there is at least one state $\rho_{ABC}\in H_A\otimes H_B\otimes H_C$ such that (\ref{m12}) is not satisfied for any given $\alpha$.

With respect to each quantum state $\rho_{ABC}\in H_A\otimes H_B\otimes H_C$, we define
a state dependent parameter $x_{\rho_{ABC}}$ such that the following equation is satisfied,
\begin{equation}\label{EQ1}
x_{\rho_{ABC}}\left(E^{y}(\rho_{A|BC})-\max\{E^{y}(\rho_{AB}),E^{y}(\rho_{AC})\}\right)
=\min\{E^{y}(\rho_{AB}),E^{y}(\rho_{AC})\},
\end{equation}
where $y$ is a positive number.
If Eq.(\ref{EQ1}) has non-zero solution $x_{\rho_{ABC}}$ for each $\rho_{ABC}$ with $y=1$, then Eq. (\ref{EQ1}) becomes Eq.(\ref{jine}), i.e., $x_{\rho_{ABC}}=\frac{1}{\mu}$.
Hereafter, we take $x_{\rho_{ABC}}=0$ for any $y$ when $E(\rho_{A|BC})=\max\{E(\rho_{AB}),E(\rho_{AC})$ and $\min\{E(\rho_{AB}),E(\rho_{AC})\}=0$.
For simplicity, we denote the set of parameters $x_{\rho_{ABC}}$ by
\begin{equation*}
X_{y}=\left\{x_{\rho_{ABC}}|x_{\rho_{ABC}}\left(E^{y}(\rho_{A|BC})
-\max\{E^{y}(\rho_{AB}),E^{y}(\rho_{AC})\}\right)
=\min\{E^{y}(\rho_{AB}),E^{y}(\rho_{AC})\}, \rho_{ABC}\in H_A\otimes H_B\otimes H_C\right\}.
\end{equation*}

\begin{theorem}\label{TH1}
A quantum entanglement measure $E$ is $\alpha$-monogamous if and only if there exists a real number $y_0>0$ such that $X_{y_0}$ is a bounded set.
\end{theorem}

{\sf [Proof]}~
Since as a measure of quantum entanglement, $E$ does not increase under local operations and classical communication, we have $E(\rho_{A|BC})\geq \max\{E(\rho_{AB}),E(\rho_{AC})\}$. Without loss of generality, we assume $E(\rho_{AB})\geq E(\rho_{AC})$ (The case for $E_{AC}\geq E_{AB}$ is similarly treated).

$(\Rightarrow)$~~ If $E$ is $\alpha$-monogamous, then $E$ satisfies (\ref{m12}) for any $\rho_{ABC}\in H_A\otimes H_B\otimes H_C$.

(Case 1) For a given state $\rho_{ABC}$, if $E(\rho_{A|BC})>E(\rho_{AB})$, we take $y=\alpha$, then $x_{\rho_{ABC}}=\frac{E^{\alpha}(\rho_{AC})}{E^{\alpha}(\rho_{A|BC})-E^{\alpha}(\rho_{AB})}.$ From (\ref{m12}) we have $x_{\rho_{ABC}}\leq1.$

(Case 2) If $E(\rho_{A|BC})=E(\rho_{AB})$, since $E^{\alpha}_{A|BC}\geq
E^{\alpha}_{AB}+E^{\alpha}_{AC}$, we have $E_{AC}=0$.
The solution of Eq. (\ref{EQ1}) is $x_{\rho_{ABC}}=0$ for any $y$, especially for $y=\alpha$.

From Case 1 and Case 2 we have $x_{\rho_{ABC}}\in[0,1]$ for each $x_{\rho_{ABC}}\in X_{\alpha}$. Therefore, $X_{\alpha}$ is a bounded set.

$(\Leftarrow)$~~ If $X_{y_0}$ is a bounded set,
then there exists a number $M>0$ such that $0\leq x_{\rho_{ABC}}\leq M$ for any $x_{\rho_{ABC}}\in X_{y_0}$.

(Case 1) If $x_{\rho_{ABC}}>0$, we have $E(\rho_{A|BC})>E(\rho_{AB})\geq E(\rho_{AC})>0$.
Let $g(\beta)=E^{\beta}(\rho_{A|BC})- E^{\beta}(\rho_{AB})-E^{\beta}(\rho_{AC})$.
We get
\begin{eqnarray*}\nonumber
g(\beta)&&=E^{\beta}(\rho_{A|BC})-
 E^{\beta}(\rho_{AB})-E^{\beta}(\rho_{AC})\\
 &&=\left(E^{y_0}(\rho_{AB})+\frac{1}{x_{\rho_{ABC}}}E^{y_0}
 (\rho_{AC})\right)^{\frac{\beta}{y_0}}-E^{\beta}(\rho_{AB})-E^{\beta}(\rho_{AC})\\
 &&=
 E^{\beta}(\rho_{AB})\left(1+\frac{1}{x_{\rho_{ABC}}}\frac{E^{y_0}(\rho_{AC})}
 {E^{y_0}(\rho_{AB})}\right)^{\frac{\beta}{y_0}}
 -E^{\beta}(\rho_{AB})-E^{\beta}(\rho_{AC})\\
  &&\geq E^{\beta}(\rho_{AB})\left(1+\frac{\beta}{x_{\rho_{ABC}}y_0}\frac{E^{y_0}(\rho_{AC})}
  {E^{y_0}(\rho_{AB})}\right)
  -E^{\beta}(\rho_{AB})-E^{\beta}(\rho_{AC})\\
 &&=E^{y_0}(\rho_{AC})\left(\frac{\beta}{x_{\rho_{ABC}}y_0}E^{\beta-y_0}(\rho_{AB})-E^{\beta-y_0}(\rho_{AC})\right),
 \end{eqnarray*}
where the second equality is due to (\ref{EQ1}) and the first inequality is due to that
$(1+t)^{n}\geq1+nt$ for $n\geq1$ and $t\geq0$.
Setting $\beta=\max\{My_0,y_0\}$, we obtain $g(\beta)\geq0$ for the given state $\rho_{ABC}$ since $0< x_{\rho_{ABC}}\leq M$ and $E(\rho_{AB})\geq E(\rho_{AC})$.

(Case 2) If $x_{\rho_{ABC}}=0$ for the quantum state $\rho_{ABC}$, one has $E(\rho_{AC})=0$.
Then $E^{\alpha}(\rho_{A|BC})\geq E^{\alpha}(\rho_{AB})+E^{\alpha}(\rho_{AC})$ for any $\alpha>0$.

In conclusion, if there exists a real number $y_0$ such that $X_{y_0}$ is a bounded set,
we take $\alpha=\max_{x_{\rho_{ABC}}>0}\{\beta\}=\max\{My_0,y_0\}$. Then $g(\alpha)\geq0$, i.e., $E^{\alpha}(\rho_{A|BC})\geq E^{\alpha}(\rho_{AB})+E^{\alpha}(\rho_{AC})$
for each $\rho_{ABC}\in H_A\otimes H_B\otimes H_C.$
\qed

In fact, from the proof of the Theorem 1, we have $g(\beta)\geq0$ for all $\beta\geq\alpha.$
Namely, if $E$ is $l$-monogamous, $E$ must be $L$-monogamous with $L\geq l$.
From Theorem 1 we can conclude that the monogamy of $E$ is determined by $X_{y}$.

As applications let us consider the concurrence and three-qubit systems. For
any arbitrary bipartite pure state $|\psi\rangle_{AB}$, the concurrence is given by
$C(|\psi\rangle_{AB})=\sqrt{1-Tr(\rho^2_{A})}$, where $\rho_A=Tr_B(|\psi\rangle_{AB}\langle\psi|)$.
We first consider the solution of Eq. (\ref{EQ1}) for three-qubit with $E=C$ and $y=2$. Any three-qubit state $|\psi\rangle_{ABC}$ can be written in
the generalized Schmidt decomposition \cite{gx},
\begin{equation}\label{pure3}
\ba{rcl}
|\psi\rangle_{ABC}&=&\lambda_0|000\rangle+
\lambda_1e^{i\varphi}|100\rangle
+\lambda_2|101\rangle
+\lambda_3|110\rangle+
\lambda_4|111\rangle,
\ea
\end{equation}
where $\lambda_i\geq0$, $i=0,...,4$, and $\sum_{i=0}^{4}\lambda_i^2=1$.
We have $C_{A|BC}=2\lambda_0\sqrt{\lambda^2_2+\lambda^2_3+\lambda^2_4},$
$C_{AB}=2\lambda_0\lambda_2$ and $C_{AC}=2\lambda_0\lambda_3.$

Denote $\lambda=\min\{\lambda_2,\lambda_3\}$. Eq. (\ref{EQ1}) becomes
$x_{|\psi\rangle_{ABC}}\lambda^2_0(\lambda^2+\lambda^2_4)=\lambda^2_0\lambda^2$. We obtain the following solutions,
\begin{equation}\label{x11}
x_{|\psi\rangle_{ABC}}=\left\{
\begin{aligned}
&~~~ 0,~~~ ~~~~~~ \lambda_0=0;\\
&~~~ 0,~~~~~~~~~  \lambda_0\not=0~~and~~ \lambda=0;\\
&\frac{\lambda^2}{\lambda^2+\lambda^2_4},~~~\lambda_0\not=0~~and~~ \lambda\not=0.
\end{aligned}
\right.
\end{equation}
We have $X_2=\{x_{|\psi\rangle_{ABC}}|x_{|\psi\rangle_{ABC}}\lambda^2_0(\lambda^2+\lambda^2_4)
=\lambda^2_0\lambda^2\}$ and $\max\{x_{|\psi\rangle_{ABC}}\}=1$.
According the proof of Theorem \ref{TH1}, we take $M=1$. Then $\alpha=\max\{2M,2\}=2$,
i.e., $C^2(|\psi\rangle_{A|BC})\geq C^2(\rho_{AB})+C^2(\rho_{AC})$ for any pure states $|\psi\rangle_{ABC}$.

Since the concurrence of any mixed state $\rho_{AB}=\sum_ip_i|\psi\rangle_i\langle\psi|$ is given by the convex roof extension, $C(\rho_{AB})=\min_{\{p_i,|\psi\rangle_i\}}\sum_ip_iC(|\psi\rangle_{i})$,
for three-qubit mixed states $\rho_{A|BC}$, we have also
$C^2(\rho_{A|BC})\geq C^2(\rho_{AB})+C^2(\rho_{AC})$.

In Theorem 1, we established a necessary and sufficient
condition for the monogamy of a quantum entanglement measure $E$.
In the following, we consider  the generalized  measures of quantum correlation beside  entanglement measure $E$, such as entanglement of assistance\cite{ga}, quantum discord\cite{85040102} and quantum
deficit\cite{85012103} and so on.
Hereafter, we  also say  $Q$ is $\alpha$-monogamous if there exists a real number $\alpha>0$
such that $Q$ satisfies (\ref{m12}) and $Q$ is non-monogamous if there is at least one state $\rho_{ABC}\in H_A\otimes H_B\otimes H_C$ such that (\ref{m12}) is not satisfied for any given $\alpha$.

Different from the concurrence, the concurrence of assistance $C_a$ fails to be $\alpha$-monogamous in general. For a tripartite state $|\psi\rangle_{ABC}$, the concurrence of assistance is defined by \cite{ca}
\begin{equation}
C_a(|\psi\rangle_{ABC})\equiv C_a(\rho_{AB})
=\max_{\{p_i,|\psi_i\rangle\}}\sum_ip_iC(|\psi_i\rangle),
\end{equation}
where the maximum takes over all possible ensemble realizations of
$\rho_{AB}=Tr_{C}(|\psi\rangle_{ABC}\langle\psi|)=\sum_i p_i |\psi_i\rangle_{AB} \langle \psi_i|$.
Consider the $2\otimes 2\otimes 2$ pure state $|\psi\rangle_{ABC}$ given in (\ref{pure3}).
One has $C_{a}(\rho_{AB})=2\lambda_0\sqrt{\lambda^2_3+\lambda^2_4}$ , $C_{a}(\rho_{AC})=2\lambda_0\sqrt{\lambda^2_2+\lambda^2_4}$
and  $C_a(|\psi\rangle_{A|BC}\rangle)=2\lambda_0\sqrt{\lambda^2_2+\lambda^2_3+\lambda^2_4}$.
In particular, when $\lambda_0\not=0,$ $\lambda_2=0$ and $\lambda_4\not=0$, i.e., $C_a(|\psi\rangle_{A|BC}\rangle)=C_{a}(\rho_{AB})>0$ and $C_{a}(\rho_{AC})>0$,
the Eq. (\ref{EQ1}) has no bounded solution for any $y$.
The set $\Gamma=\left\{\beta\,(0<\beta<+\infty)|C^{\beta}_a(\rho_{A|BC})\geq C_a^{\beta}(\rho_{AB})+C_a^{\beta}(\rho_{AC})~for~ all~\rho\in2\otimes 2\otimes 2\right\}$ is just an empty set.

\begin{theorem}\label{TH2}
For a given  measure of quantum correlation $Q$ with $Q(\rho_{A|BC})\geq\max\{Q(\rho_{AB}),Q(\rho_{AC})\}$, if there exists a state $\rho_{ABC}\in H_A\otimes H_B\otimes H_C$ such that
$Q(\rho_{A|BC})=\max\{Q(\rho_{AB}),Q(\rho_{AC})\}$ and
$\min\{Q(\rho_{AB}),Q(\rho_{AC})\}>0$, then $Q$ is non-monogamous for the quantum system $ABC$.
\end{theorem}

{\sf [Proof]}
If $\rho_{ABC}$ satisfies $Q(\rho_{A|BC})=\max\{Q(\rho_{AB}),Q(\rho_{AC})\}$ and $\min\{Q(\rho_{AB}),Q(\rho_{AC})\}>0$, then
$Q^{\alpha}(\rho_{A|BC})< Q^{\alpha}(\rho_{AB})+Q^{\alpha}(\rho_{AC})$
for any $\alpha>0$.

 Therefore, there is no positive number $\alpha$ such that
(\ref{m12}) is satisfied. \qed

\noindent{\it Example 1}
Let us consider the $2\otimes 2\otimes 3$ pure state
\begin{equation}\label{ex}
|\psi\rangle_{ABC}
=\frac{1}{\sqrt{3}}(|000\rangle+|111\rangle+|\varphi^{+}\rangle|2\rangle),
\end{equation}
where $|\varphi^{+}\rangle=\frac{1}{\sqrt{2}}(|01\rangle+(|10\rangle))$ is one of the Bell states. One has \cite{zhu3},
$C_{a}(|\psi\rangle_{A|BC}\rangle)=C_{a}(\rho_{AB})=1$ and $C_{a}(\rho_{AC})=\frac{2\sqrt{2}}{3}$.
According to Theorem \ref{TH2}, $C_a$ is non-monogamous for the $2\otimes 2\otimes 3$ quantum system.

The above results show that the concurrence of assistance $C_a$ is non-monogamous in general for the $2\otimes 2\otimes n$ ($n\in\{2,3\}$) quantum systems. Nevertheless, for particular
states, $C_a$ may be $\alpha$-monogamous. In \cite{zhu2} it is shown that $C_a$ satisfies
the monogamy relation, $C_a^{\beta}(|\psi\rangle_{ABC})\geq C_a^{\beta}(\rho_{AB})+C_a^{\beta}(\rho_{AC})$,
for the three-qubit W-class states $|\psi\rangle_{ABC}$ with $\beta\geq2$.
Therefore, with respect to a given measure of quantum  correlation $Q$, there may be exist $\alpha$,  which $Q$ satisfied $Q^{\alpha}(\rho_{A|BC})\geq Q^{\alpha}(\rho_{AB})+Q^{\alpha}(\rho_{AC})$ for specifically  states $\rho_{ABC}$.
We denote
\begin{equation*}
X_{y,\Omega}=\left\{x_{\rho_{ABC}}|x_{\rho_{ABC}}\left(Q^{y}(\rho_{A|BC})
-\max\{Q^{y}(\rho_{AB}),Q^{y}(\rho_{AC})\}\right)
=\min\{Q^{y}(\rho_{AB}),Q^{y}(\rho_{AC})\},
where~\rho_{ABC}\in \Omega\right\}.
\end{equation*}
Together with the Theorem \ref{TH1} and Theorem \ref{TH2},
we have the following result.

\begin{corollary}\label{c3}
Given a given  measure of quantum correlation $Q$  and a nonempty set of quantum states $\Omega=\left\{\rho_{ABC}|Q(\rho_{A|BC})\geq\max\{Q(\rho_{AB}),Q(\rho_{AC})\}\right\}$.
$Q$ is $\alpha-$monogamous for $\rho_{ABC}\in\Omega$ if and only if there exists a real number $y_0$ such that $X_{y_0,\Omega}$ is a bounded set.
\end{corollary}

The concurrence of assistance $C_a$  is an entanglement monotone for  $2\otimes 2\otimes n(n\geq2)$ pure states\cite{72042329}, i,e $C_{a}(|\psi\rangle_{A|BC})\geq\max\{C_a(\rho_{AB}),C_a(\rho_{AC})\}$ for quantum state $|\psi\rangle_{ABC}\in2\otimes 2\otimes n(n\geq2)$.
Then we can study the monogamy of  $C_a$ for $2\otimes 2\otimes n(n\geq2)$ pure states by the corollary \ref{c3}.

\noindent{\it Example 2}
Consider the three-qubit W-class states $|\psi\rangle_{ABC}$ \cite{zhu2},
\begin{equation}\label{W}
|\psi\rangle_{ABC}=b_0|000\rangle+b_1|100\rangle+b_2|010\rangle+b_3|001\rangle,
\end{equation}
where $\sum_{i=0}^{3}|b_i|^2=1$. Denote $\Omega=\left\{\rho_{ABC}|\rho_{ABC}=|\psi\rangle_{ABC}\langle\psi|\right\}$.
One has $C_a(|\psi\rangle_{ABC})=2|b_1|\sqrt{|b_2|^2+|b_3|^2}$, $C_a(\rho_{AB})=2|b_1||b_2|$
and $C_a(\rho_{AC})=2|b_1||b_3|$.
From Theorem 1 and Corollary \ref{c3} we have
\begin{equation*}
x_{|\psi\rangle_{ABC}}=\left\{
\begin{aligned}
&~~~ 0,~~~ ~~~~~~ b_1 b_2b_3=0;\\
&~~~1,~~~~~~~~~b_1b_2b_3\not=0,
\end{aligned}
\right.
\end{equation*}
then $X_{2,\Omega_{0}}=\{0,1\}$. Hence $C_a$ is $2$-monogamous for $\rho_{ABC}\in\Omega$.

Theorem \ref{TH1} and Corollary \ref{c3} provide the monogamy of $Q$ based on value of $x_{\rho_{ABC}}\in X_y$ for given $y$. In the following, we provide the $y$ independent  monogamy with respect to a given quantum state $\rho_{ABC}$.
For convenience, we denote $\Lambda=\{\rho_{ABC}|Q(\rho_{A|BC})>Q(\rho_{AB})\geq Q(\rho_{AC})>0~or~Q(\rho_{A|BC})>Q(\rho_{AC})\geq Q(\rho_{AB})>0\}$.

\begin{theorem}\label{TH3}
For a given  measure of quantum correlation $Q$ and a quantum state $\rho_{ABC}\in\Lambda$,
$Q$ is $\alpha$-monogamous according to (\ref{m12}) with $\alpha=\log_{b}2$, where $b=\min\left\{\frac{Q(\rho_{A|BC})}{Q(\rho_{AC})},\frac{Q(\rho_{A|BC})}{Q(\rho_{AB})}\right\}$.
\end{theorem}

{\sf [Proof]}~Without loss of generality,
we assume $Q(\rho_{A|BC})>Q(\rho_{AB})\geq Q(\rho_{AC})>0$.
Set $t_1=\frac{Q(\rho_{AB})}{Q(\rho_{A|BC})}$, $t_2=\frac{Q(\rho_{AC})}{Q(\rho_{A|BC})}$
and  $k=\frac{Q(\rho_{AB})}{Q(\rho_{AC})}$.
It is obvious that $t_1\geq t_2$, $k\geq1$ and
$1=t^y_1+\frac{1}{x}t^y_2$,
where $x$ is the solution of Eq. (\ref{EQ1}) with a given $y>0$ for $\rho_{ABC}$.
Let $\alpha$ satisfy $(1+\frac{1}{k^yx})^{\frac{\alpha}{y}}=2$, i.e.,
$\alpha=\log_{\frac{Q(\rho_{A|BC})}{Q(\rho_{AB})}}2$.
Next, we prove that $g(\alpha)=Q^{\alpha}(\rho_{A|BC})(1-t_1^{\alpha}-t_2^{\alpha})\geq 0$
by proving $(1-t_1^{\alpha}-t_2^{\alpha})\geq 0$,
\begin{eqnarray*}\nonumber
1-t_1^{\alpha}-t_2^{\alpha}
 &&=(t^y_1+\frac{1}{x}t^y_2)^{\frac{\alpha}{y}}-t_1^{\alpha}-t_2^{\alpha}\\
 &&=
 (k^y+\frac{1}{x})^{\frac{\alpha}{y}}t_2^{\alpha}-k^{\alpha}t_2^{\alpha}
 -t_2^{\alpha}\\
  &&=k^{\alpha}
  \left((1+\frac{1}{k^yx})^{\frac{\alpha}{y}}-1-k^{-\alpha}\right)t_2^{\alpha}\\
 &&=(k^{\alpha}-1)t_2^{\alpha}\geq0,
 \end{eqnarray*}
where the first equation is due to $1=t^y_1+\frac{1}{x}t^y_2$ and the inequality is due to
$k\geq1$ and $\alpha>0$. The case for $Q(\rho_{A|BC})>Q(\rho_{AC})\geq Q(\rho_{AB})>0$
is similarly proved. Therefore, $Q$ is $\log_{b}2$-monogamous for the states $\rho_{ABC}$. \qed

The entanglement cost $E_C$ of a bipartite state $\rho_{AB}$ is equal to the regularized entanglement of formation $E_f$ \cite{EOF}, namely \cite{m1},
\begin{equation}
E_C(\rho_{AB})=\lim_{n\to\infty}\frac{1}{n}E_{f}(\rho^{\otimes n}_{AB}),
\end{equation}
where the entanglement of formation $E_{f}$ is defined by
$E_{f}(\rho_{AB})=\min_{\{p_i,|\psi_i\rangle\}}\sum_ip_iS(Tr_B[|\psi_i\rangle\langle\psi_i|]),$
with $S(\rho)$ is the von Neumann entropy of density operator $\rho$
and the minimum is taken over all ensembles $\{p_i,|\psi_i\rangle\}$ satisfying $\sum_ip_i|\psi_i\rangle\langle\psi_i|=\rho_{AB}$.
It is worthwhile noting that the inequality (\ref{m11}) does not hold for the entanglement cost $E_C$ \cite{m1}.

To illustrate the Theorem \ref{TH3} we consider the purification of the totally antisymmetric two-qutrit state,
\begin{equation}\label{5}
|\psi\rangle_{ABC}=\frac{1}{\sqrt{6}}(|123\rangle
-|132\rangle+|231\rangle-|213\rangle+|321\rangle-|321\rangle).
\end{equation}
One has $E_C(|\psi\rangle_{A|BC})=\log_{2}3$ and  $E_C(\rho_{AB})=E_C(\rho_{AB})=1$ \cite{m1}.
According to Theorem \ref{TH3}, we have  $b=\min\left\{\frac{E_C(|\psi\rangle_{A|BC})}{E_C(\rho_{AC})},\frac{E_C(|\psi\rangle_{A|BC})}
{E_C(\rho_{AB})}\right\}=\log_{2}3$. Then $\alpha=\log_{(\log_{2}3)}2\approx1.51.$
Therefore, $E_C$ is $\alpha$-monogamous for $|\psi\rangle_{ABC}$ for $\alpha\geq1.51$.
Denote $f(\alpha)=E^{\alpha}_C(|\psi\rangle_{A|BC})-E^{\alpha}_C(\rho_{AB})
-E^{\alpha}_C(\rho_{AC})$. We have $f(\alpha)\geq0$ for $\alpha\geq1.51$,
see the Fig. 1.
\begin{figure}[htpb]
\renewcommand{\figurename}{Fig.}
\centering
\includegraphics[width=8.5cm]{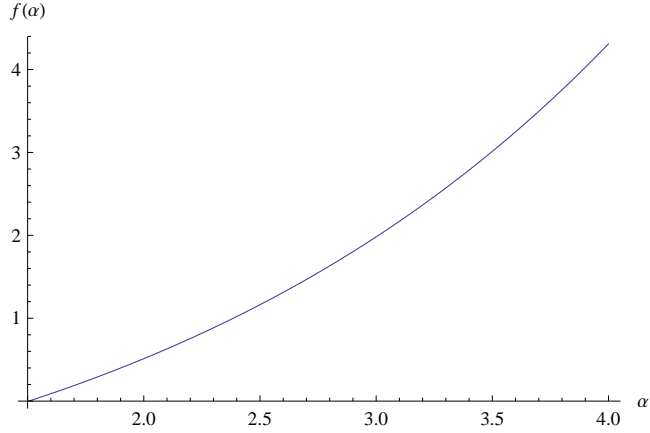}
\caption{{\small $f(\alpha)$ as a function of $\alpha$ ($\alpha\geq1.51$) for $|\psi\rangle_{ABC}$.}}
\end{figure}

On the other hand, for the state $|\psi\rangle_{ABC}$ in (\ref{5}), using the Theorem 1 and  Corollary \ref{c3}, we have $X_{y,|\psi\rangle_{ABC}}=\{\frac{1}{(\log_{2}3)^y-1}\}$ for given $y>0$. Then $E_C$ is $\beta$-monogamous for $|\psi\rangle_{ABC}$ with $\beta=\max\{\frac{1}{(\log_{2}3)^y-1}y,y\}$.
Obviously, the optimal value of $\beta$ is obtained at the intersection
of $z_1(y)=\frac{1}{(\log_{2}3)^y-1}y$ and $z_2(y)=y$ with $y>0$, i.e.,
$\min_{y>0}\{\beta\}=\log_{(\log_{2}3)}2$, see Fig. 2.
\begin{figure}[htpb]
\renewcommand{\figurename}{Fig.}
\centering
\includegraphics[width=8.5cm]{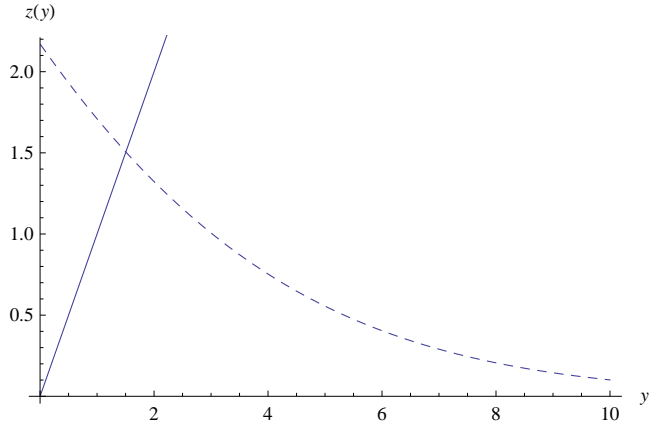}
\caption{{\small The dished line is for $z_1(y)$ and the solid line is for $z_2(y)$.}}
\end{figure}

In the proof of the Theorem \ref{TH3}, if we take $1<c\leq \min\left\{\frac{Q(\rho_{A|BC})}{Q(\rho_{AC})},\frac{Q(\rho_{A|BC})}{Q(\rho_{AB})}\right\}$, then we have that $Q$ is $\log_{c}2$-monogamous for the given quantum state.
For example, let $c=\frac{3}{2}\in(1,\log_{(\log_{2}3)}2]$ for the state (\ref{5}).
Then $E_C$ is $\alpha$-monogamous for $|\psi\rangle_{ABC}$ with $\alpha=\log_{\frac{3}{2}}2\approx1.71$. More generally,
for some states $\rho_{ABC}\in\Lambda$, from the proof of Theorem \ref{TH3} if $T =\left\{\frac{Q(\rho_{A|BC})}{Q(\rho_{AC})},\frac{Q(\rho_{A|BC})}{Q(\rho_{AB})}
\big|\rho_{ABC}\right\}$ has lower bound $c$ which is strictly greater than $1$, $Q$ is $\log_{c}2$-monogamous for those states $\rho_{ABC}$.

\section{Conclusions and Remarks}

We have addressed in general the question of whether entanglement measures $E$ are
monogamous in the sense of (\ref{m12}) introduced in Ref. \cite{G1} for tripartite quantum systems. We have presented the equations (\ref{EQ1}) and shown that
the monogamy of a entanglement measure $E$ only depends on the boundedness of the solutions for the equations. Both sufficient and necessary conditions for $E$ being monogamous have been explicitly derived. Meanwhile, the monogamy conditions for a  measure of  quantum correlation $Q$  have been also obtained with respect to certain subsets of quantum states. According to the duality, our approach may be also applied to study the polygamy of entanglement measures $E$.

\bigskip
\noindent{\bf Acknowledgments}\, \,
This work is supported by the National Natural Science Foundation
of China under grant Nos. 12075159 and 12171044, Beijing Natural Science Foundation (Z190005), and the Academician Innovation Platform of Hainan Province.

\end{document}